\documentclass[10pt]{article}
\usepackage{float}
\usepackage[symbol]{footmisc}
\usepackage[superscript,sort]{cite}
\usepackage{array}
\usepackage[pdftex,colorlinks=true,linkcolor=black,citecolor=black,urlcolor=black,bookmarks=true,breaklinks=true]{hyperref}
\usepackage[pdftex]{graphicx}
\usepackage[usenames,pdftex]{color}
\definecolor{Red}{rgb}{1.0,0.0,0.0}
\usepackage{amsmath,amssymb}

\usepackage{amsthm,amscd,amsxtra,amsfonts,amsmath,amssymb,multirow}
\usepackage{wrapfig}
\usepackage[footnotesize]{caption}
\usepackage[tiny,compact]{titlesec}
\usepackage{ctable}
\usepackage{helvet}

\usepackage{xcolor,colortbl}
\usepackage{subfigure}
\usepackage{tikz}
\usetikzlibrary{shapes,arrows}
\usepackage[T1]{fontenc}

\titlespacing*{\section}{0pt}{*0}{*0}
\titlespacing*{\subsection}{0pt}{*0}{*0}
\titlespacing*{\subsubsection}{0pt}{*0}{*0} 
\titlespacing{\paragraph}{0pt}{*0}{*1}

\definecolor{MyPurple}{rgb}{1,0,1}

\newcommand{\beq}[1]{\begin{equation} \label{#1}}
\newcommand{\eeq}{\end{equation}}
\newcommand{\barray}{\begin{array}{ll}}
\newcommand{\earray}{\end{array}}

\begin{document}
\pagenumbering{roman}

\clearpage \pagebreak \setcounter{page}{1}
\renewcommand{\thepage}{{\arabic{page}}}

\title{Capturing protein multiscale thermal fluctuations
}

\author{
Kristopher Opron$^1$,
Kelin Xia$^2$  and
Guo-Wei Wei$^{1,2,3}$ \footnote{ Address correspondences  to Guo-Wei Wei. E-mail:wei@math.msu.edu}\\
$^1$  Department of Biochemistry and Molecular Biology\\
Michigan State University, MI 48824, USA \\
$^2$ Department of Mathematics \\
Michigan State University, MI 48824, USA\\
$^3$ Department of Electrical and Computer Engineering \\
Michigan State University, MI 48824, USA \\
}

\date{\today}
\maketitle

\begin{abstract}

Existing elastic network models are typically parametrized at a given cutoff distance and often fail to properly predict the thermal fluctuation of many macromolecules  that involve  multiple characteristic length scales. We  introduce a multiscale flexibility-rigidity index (mFRI) method to  resolve this problem.  The proposed mFRI utilizes two or three correlation kernels parametrized at different length scales to  capture  protein interactions at corresponding scales. It is about 20\% more accurate than  the Gaussian network model (GNM) in the B-factor prediction of a set of 364 proteins. Additionally, the present method is able to delivery accurate predictions for multiscale macromolecules that fail GNM.    Finally, or a protein of $N$ residues,  mFRI is of linear scaling (${\cal O}(N)$) in computational complexity, in contrast to the order of ${\cal O}(N^3)$  for GNM.

\end{abstract}
\maketitle


Proteins  are among the most essential biomolecules for  life.  Many protein functions, such as structure support,   catalyzing chemical reactions,  and allosteric regulation are strongly correlated  to protein flexibility \cite{Frauenfelder:1991}. Protein flexibility is   an intrinsic property of proteins  and  can be measured directly or indirectly by many experimental approaches, such as  X-ray crystallography,  nuclear magnetic resonance (NMR) and single-molecule force experiments \cite{Dudko:2006}.  Theoretically, protein flexibility can be computed by     normal mode analysis (NMA)  \cite{Go:1983,Tasumi:1982,Brooks:1983,Levitt:1985}, graph theory \cite{Jacobs:2001}, { rotation translation blocks (RTB) method \cite{tama:2000,Demerdash:2012},} and elastic network model (ENM) \cite{Bahar:1997,Bahar:1998,Atilgan:2001,Hinsen:1998,Tama:2001,LiGH:2002},  including    Gaussian network model (GNM)   \cite{Bahar:1997,Bahar:1998}  and anisotropic network model (ANM) \cite{Atilgan:2001}.  A common feature of the above mentioned time-independent methods is that they resort to the matrix diagonalization procedure. The computational complexity of  the matrix diagonalization is typically of the order of ${\cal O}(N^3)$, where $N$ is the number of elements in the matrix. Such a computational complexity calls for new efficient strategies for the flexibility analysis of large biomolcules.

It is well known that NMA and GNM do not work well for many macromolecules.  Park et al. had collected three sets of structures to test performance of NMA and GNM methods \cite{JKPark:2013}.  It was found that   both methods fail to work and deliver negative correlation coefficients for many structures  \cite{JKPark:2013}.  The mean correlation coefficients (MCCs) for the B-factor prediction of small-sized, medium-sized and large-sized  sets of structures  are about  0.480, 0.482 and 0.494 for NMA, respectively \cite{JKPark:2013,Opron:2014}. The GNM preforms slightly better, with the mean  correlation coefficients of 0.541, 0.550 and 0.529 for the above test sets  \cite{JKPark:2013,Opron:2014}. Obviously, there is a pressing need to develop innovative approaches for biomolecular flexibility analysis.

Recently, we have proposed a few matrix-decomposition-free methods for flexibility analysis, including  molecular nonlinear dynamics \cite{KLXia:2013b},  stochastic dynamics \cite{KLXia:2013f} and flexibility-rigidity index  (FRI)  \cite{KLXia:2013d,Opron:2014}. Among them,  flexibility-rigidity index  (FRI) has been introduced to evaluate protein flexibility and rigidity.  The fundamental assumptions of the FRI method are as follows. Protein functions, such as flexibility, rigidity, and energy, are fully determined by the structure of the protein and its environment, and the protein structure is in turn   determined by   the relavent interactions. Therefore, whenever the protein structure is available, there is no need to analyze protein flexibility and rigidity by tracing back to the protein interaction Hamiltonian. Consequently, the FRI bypasses the  ${\cal O}(N^3)$ matrix diagonalization. Our initial FRI  \cite{KLXia:2013d} has the computational complexity of of ${\cal O}(N^2)$ and our fast FRI (fFRI) \cite{Opron:2014} based on a cell lists algorithm \cite{Allen:1987} is of ${\cal O}(N)$. The FRI and the fFRI have been extensively validated by a set of 365 proteins for parametrization, accuracy and reliability. The parameter free fFRI is about ten percent  more accurate than the GNM on the 365 protein test set and is orders of magnitude faster than GNM on a set of 44 proteins. FRI is able to predict the B-factors of an HIV virus capsid (313 236 residues) in less than 30 seconds on a single-core  processor, which would require   GNM more than 120 years  to accomplish if the computer memory is not a problem \cite{Opron:2014}.
{See the supplementary material for detail.}

Nevertheless, there are structures for which FRI does not work either. In fact, for those structures that fail NMA and GNM are  likely to be difficult for FRI as well. One such structure is pictured in Figure \ref{calmod1} where the GNM method fails to predict the high flexibility of a hinge region in calmodulin with any cutoff distance. There are a number of reasons for this and other types of failure. Crystal environment, solvent type, co-factors,  data collection conditions, and  structural refinement procedures are well-known causes \cite{Kundu:2002,Kondrashov:2007,Hinsen:2008,GSong:2007}.

However, there is one more important cause that has not been discussed in the literature to our best knowledge, namely,  multiple characteristic length scales in a single protein structure. Indeed, contrary to small molecules,  macromolecular interactions have a wide variety  of characteristic  length scales, ranging from  covalent bond, hydrogen bond, wan der Waals bond,  residue,  alpha helix  and  beta sheet,  domain  and protein  scales. Protein flexibility is intrinsically associated with protein interactions, and thus must have a multiscale trait as well. When  GNM or FRI method is parametrized at a given cutoff or scale parameter, it captures only a subset  of the characteristic length scales but inevitably misses other characteristic length scales of the protein. Consequently, none of them is able to provide an accurate B-factor prediction.

 %
%


Multiscale flexibility-rigidity index (mFRI) is constructed to capture the multiscale collective motions of macromolecules.  We utilize multiple correlation kernels, with each kernel being  parametrized at specific scale to  characterize the multiscale  flexibility  of  macromolecules.  The $n$th flexibility index of the $i$th (coarse-grained) particle is given by
\begin{eqnarray}\label{eq:flexibility39}
 f^{n}_i & = & \frac{1}{\sum_{j=1}^N w^{n}_{j} \Phi^{n}( \|{\bf r}_i - {\bf  r}_j \|;\eta^{n}_{j} )},
 \end{eqnarray}
where  $w^{n}_{j}$ is an atomic type dependent  parameter, $\Phi^{n}( \|{\bf r}_i - {\bf  r}_j \|;\eta^{n}_{j}) $ is a correlation kernel  and $\eta^{n}_{j}$ is a scale parameter. Here ${\bf r}_i$ and ${\bf r}_j$ are the coordinates for $i$th and $j$th particles, respectively.
We seek the minimization of the form
\begin{eqnarray}\label{eq:regression29}
{\rm Min}_{a^{n},b} \left\{ \sum_i \left| \sum_{n}a^n f^{n}_i + b-B^e_i\right|^2\right\}
\end{eqnarray}
where $\{B^e_i\}$ are the experimental B-factors.
We use   generalized exponential  kernels \cite{KLXia:2013d,Opron:2014}
\begin{eqnarray}\label{eq:couple_matrix19}
\Phi^n(\|{\bf r} - {\bf r}_j \|;\eta^n_{j}) =    e^{-\left(\|{\bf r} - {\bf r}_j \|/\eta^n_{j}\right)^\kappa},    \quad \kappa >0
\end{eqnarray}
and  generalized Lorentz kernels
\begin{eqnarray}\label{eq:couple_matrix29}
 \Phi^n(\|{\bf r} - {\bf r}_j \|;\eta^n_{j}) = \frac{1}{1+ \left( \|{\bf r} - {\bf r}_j \|/\eta^n_{j}\right)^{\upsilon}},  \quad  \upsilon >0.
 \end{eqnarray}
  In principle, all parameters can be optimized. For simplicity and computational efficiency, we only determine $\{a^n\}$ and $b$ in the above  minimization process. In this work,  we limit the number of kernels to at most there and set $w_j^n=1$. Both generalized exponential kernels and generalized Lorentz kernels are employed.   More detailed description of the mFRI is given in the supplementary material.


\begin{figure}[H]
\begin{tabular}{c}
\includegraphics[width=0.6\textwidth]{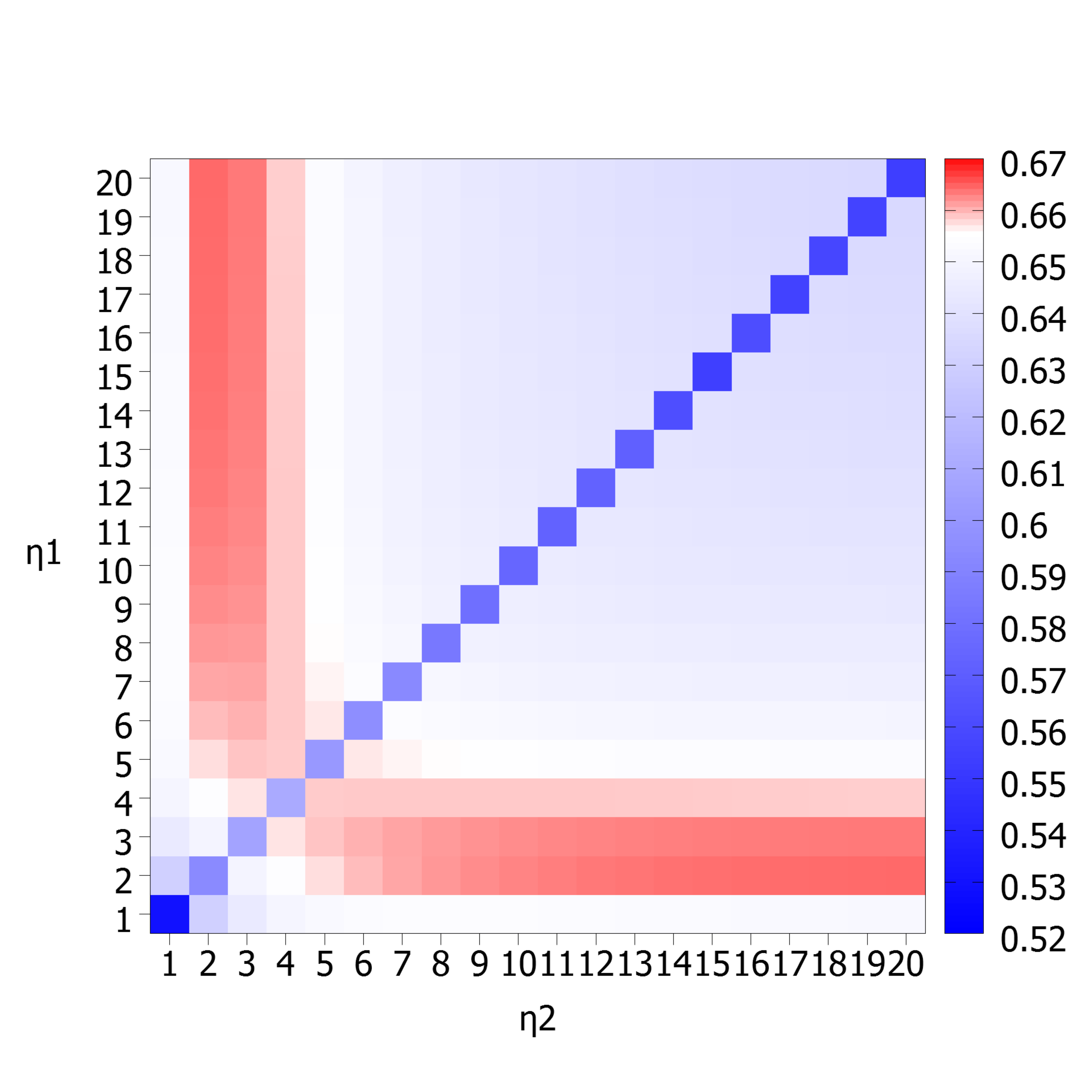}
\end{tabular}
\caption{An illustration of multiscale behavior in protein flexibility analysis. Two Lorentz kernels ($\upsilon=3$) are used. Their scale values $\eta$ values are listed along the horizontal and vertical axes. The mean correlation coefficient   value for B-factor prediction on a set of 364 proteins is shown in each cell of the matrix and color coded for convenience with red representing the highest correlation coefficients and green the lowest. Obvious, the combination of a relatively small-scale kernel and a relatively large-scale kernel delivers best prediction, which indicates the importance of incorporating multiscale in protein flexibility analysis.}
\label{2kmatrix}
\end{figure}


To understand the multiscale behavior of flexibility analysis, we consider a test set containing 364 protein structures whose  Protein Data Bank (PDB) identities are listed in the literature \cite{Opron:2014} and it contains test sets used in GNM studies \cite{JKPark:2013}. This test set omits one structure present in previous FRI studies (PDB ID: 1AGN) due to unrealistic B-factor data. Our goal is to examine how an additional kernel with a large length scale impacts the flexibility analysis. To this end, we consider two smooth  Lorentz type of kernels with $\upsilon=3$. We explore a number of scale combinations as shown in Fig. \ref{2kmatrix}, which plots  the MCC values for B-factor prediction on the set of 364 structures. The low MCC values on the diagonal line indicate that two-scale methods  are always better than a single scale one. The best results are  achieved at the combination of a  relatively small-scale  kernel and a relatively large-scale kernel.  This behavior proves  the importance of incorporating multiscale in the biomolecular flexibility analysis.  The best  MCC  for the test set is 0.67, which is about 20\% better than the best GNM prediction and about 6\% improvement over our single scale FRI approach.

The improvement in the MCC for B-factor prediction on a set of 364 proteins discussed above  obscures the fact that  the proposed multiscale method is able to captures the multiscale behaviors in  many  structures that fail the original FRI and GNM. In the rest of this paper, we demonstrate utility of the proposed multiscale method by  a few case studies.   A three-scale FRI is employed.



Protein hinge regions have been shown to be correlated with active sites and catalysis in enzymes. Flexibility has a major role in specificity of binding of a protein to other proteins, nucleic acids or other molecules. An active site or docking region that is more flexible will accommodate more varied substrates or partners while more rigid domains are more specific. Protein hinges are also found separating large domains of proteins. In this context, the hinges can be very important for protein conformational changes. The protein featured in this section, calmodulin, is a good example of a hinge that affects both structure and function.

The central region of calmodulin shown in Figure \ref{calmod1} is a long $\alpha$-helix which is unwound or kinked at the middle when no calcium is bound to the two distall metal coordinating domains. In both forms, with or without calcium bound, this helix retains a large degree of flexibility based on B-factor values from the PDB files (1CLL and 1CFD).

Many tools exist for the prediction and analysis of hinges in proteins using bioinformatics \cite{hingeatlas}, graph theory \cite{hingeprot,stonehinge,flexprot} and energetics \cite{flexoracle}.  The proposed mFRI has capabilities  similar to those in these tools. The mFRI can be used to predict hinge regions by regions of high FRI values or predicted B-values.

A comparison of   mFRI method and GNM for the B-factor prediction of calcium-bound calmodulin is displayed in Figure  \ref{calmod1}.  B-factor prediction by single kernel FRI and GNM is unable to accurately predict the hinge region in the middle of the protein with any parameter. Two- and three-kernel based mFRI methods, on the other hand, are much more accurate in the hinge region. As more kernels are added, the accuracy can be seen to grow but sufficient accuracy is achieved at three kernels.

{We have shown in our supplementary material that a similarly good B-factor prediction for  calmodulin type of structures can be achieved by the original FRI method if the crystal effect is taken into consideration. This result suggests that the proposed mFRI method may be able to take care some crystal effects.}

\newcolumntype{a}{>{\columncolor{Gray}}c}
\newcolumntype{b}{>{\columncolor{white}}c}

\begin{figure}
 \begin{center}
\begin{tabular}{cc}
\includegraphics[width=0.6\textwidth]{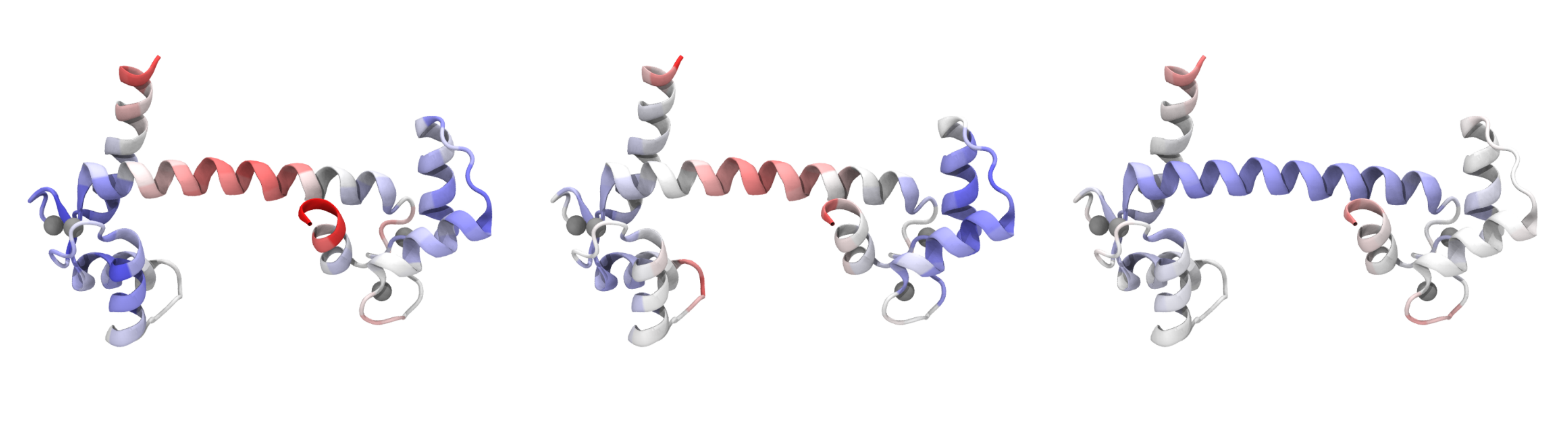}
\end{tabular}
\begin{tabular}{c}
\includegraphics[width=0.6\textwidth]{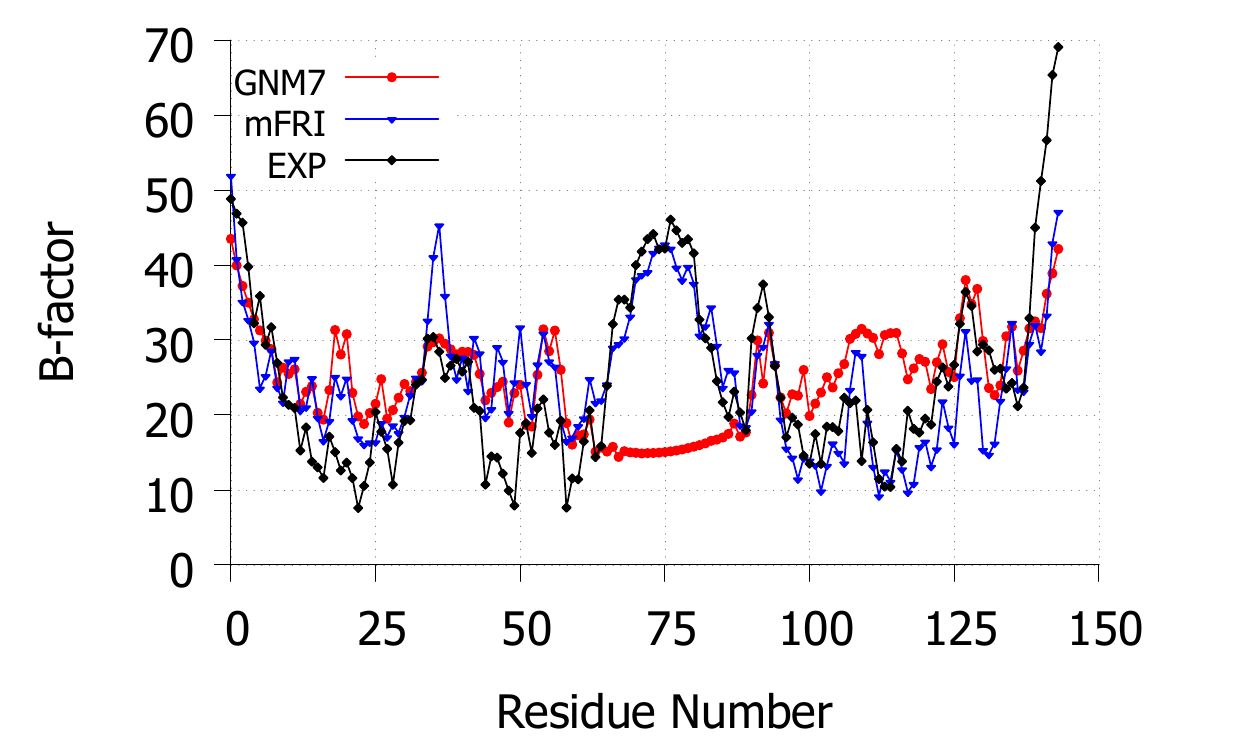}
\end{tabular}
 \end{center}
\caption{Top, the structure of calmodulin (PDB ID: 1CLL) visualized in VMD \cite{VMD} and colored by experimental B-factors (left), mFRI predicted B-factors (middle) and  GNM predicted B-factors (right)with red representing the most flexible regions. Bottom,  the experimental  and predicted B-factor values plotted per residue. The GNM7 is for the GNM method   with a cutoff distance of 7\AA. Clearly GNM misses the flexible hinge region. The mFRI is parametrized at   $\upsilon^1=3$,  $\eta^1=3$\AA, $\upsilon^2=3$,  $\eta^2=7$\AA, $\kappa^3=1$, and  $\eta^3=15$\AA.}
\label{calmod1}
\end{figure}


\begin{figure}[H]
\begin{center}
\begin{tabular}{c}
\includegraphics[width=0.6\textwidth]{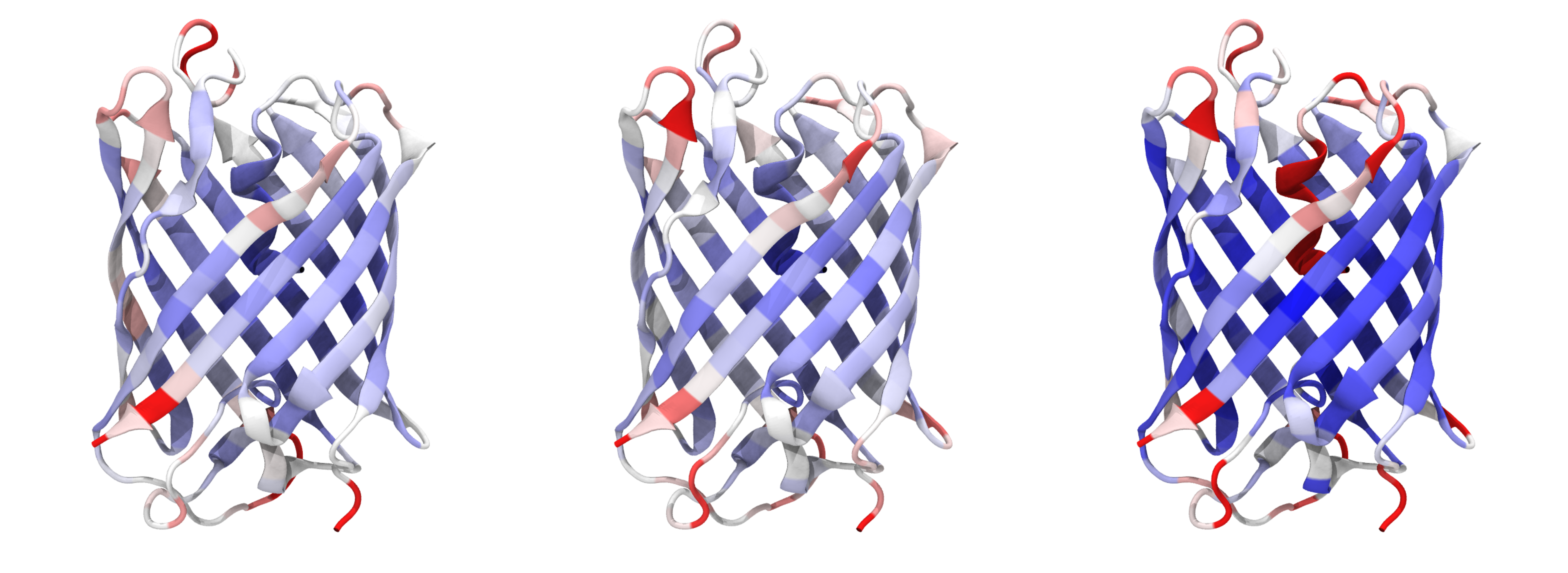}
\end{tabular}
\begin{tabular}{c}
\includegraphics[width=0.6\textwidth]{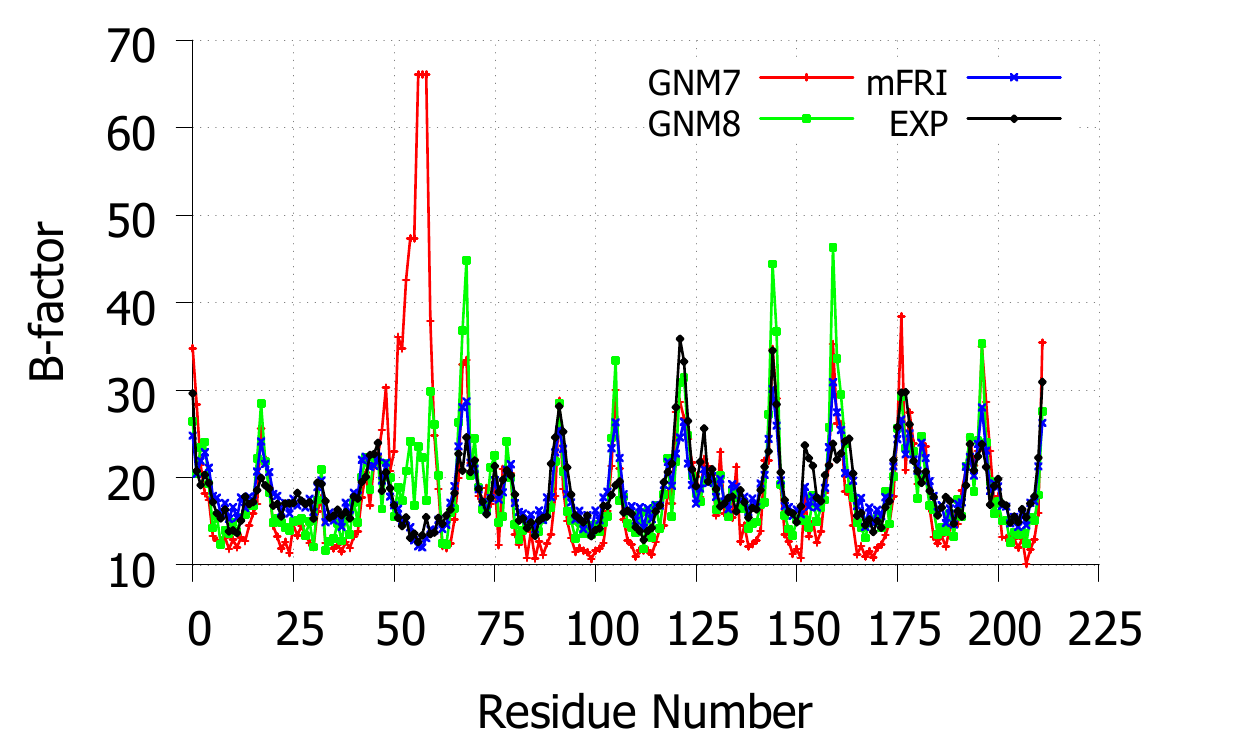}
\end{tabular}
\end{center}
\caption{ Top, a visual comparison of experimental B-factors (left), mFRI predicted B-factors (middle) and GNM predicted B-factors (right) for the engineered teal flourescent protein, mTFP1 (PDB ID:2HQK). Bottom, The experimental  and predicted B-factor values plotted per residue. The GNM naming convention indicated the cutoff used for the GNM method in \AA, i.e., GNM7 is the GNM method with a cutoff of 7\AA, etc.
}
\label{2HQK}
\end{figure}

Cyan fluorescent protein (CFP), shown in Figure \ref{2HQK}, is a homolog of the famous green flourescent protein (GFP). Isolated from the crystal jellyfish in the 1990s \cite{Shimomura1962}, GFP enabled a revolution in biochemistry by allowing the tagging and tracking of a wide range of molecules. CFP was found later in Anthozoa species which have turned out to be a good source of fluorescent proteins with varied emission spectra \cite{Matz1999}. In this example we examine the flexibility of an engineered CFP \cite{Ai2006} (PDB ID: 2HQK), mTFP1. It is clear in Figure \ref{2HQK} that GNM B-factor predictions contain a large error around residues 50-60 which is very pronounced at the recommended cutoff of 7 \AA~ and is still somewhat problematic when the cutoff is changed to 8 \AA. mFRI on the other hand has no issue with this particular region. Upon further inspection, it is clear that the offending region is the small, alpha-helical region suspended in the center of the beta-barrel. It is not surprising that this sort of configuration would be highly cutoff dependent in a scheme such as GNM, which has hard cutoffs for connectivity. It would appear that this structure is dominated by short range interaction but the region of residues 50-60 is affected to a large degree by mid-range interactions, i.e., there are at least two important scales of interaction in this case. It follows then that mFRI, which has kernels to capture short- and mid-range interactions, would perform better than GNM7 or GNM8 methods alone in B-factor predictions, Figure \ref{2HQK}, which is exactly what we see from the results.


\begin{figure}[H]
\begin{center}
\begin{tabular}{c}
\includegraphics[width=0.6\textwidth]{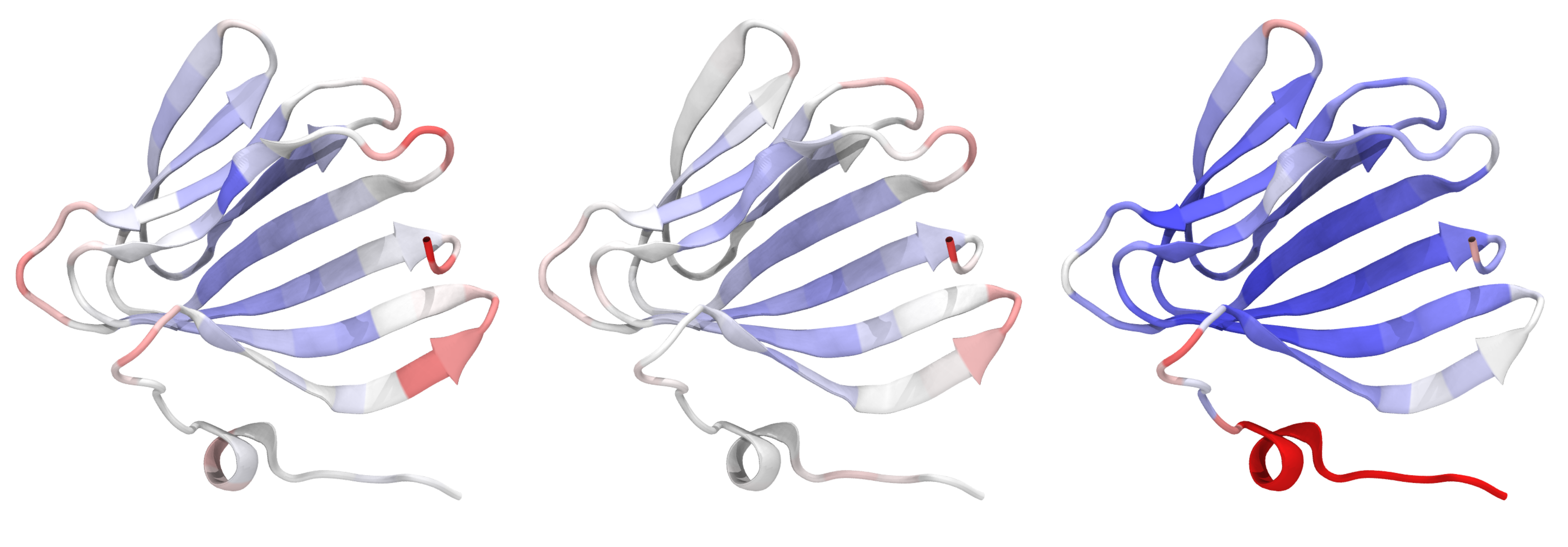}
\end{tabular}
\begin{tabular}{c}
\includegraphics[width=0.6\textwidth]{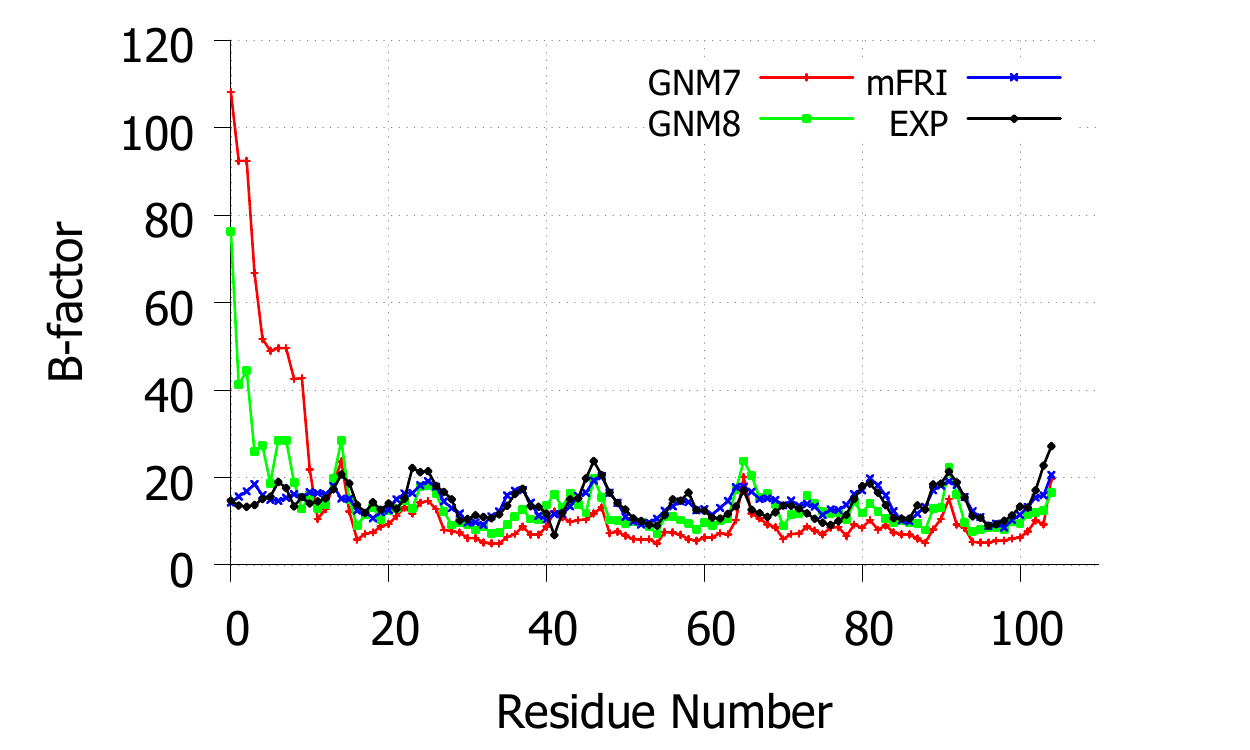}
\end{tabular}
\end{center}
\caption{ Top, a visual comparison of experimental B-factors (left), mFRI predicted B-factors (middle) and GNM predicted B-factors (right) for the engineered teal flourescent protein, mTFP1 (PDB ID:1V70). Bottom, The experimental and predicted B-factor values plotted per residue.
}
\label{1v70}
\end{figure}

A similar situation exists with the structure 1V70, a probable antibiotic synthesis protein, which is shown in Figure \ref{1v70}. As in the last example, the problematic portion for B-factor prediction comes at the end of a protein chain. In this case there is an overestimation of flexibility for residues 1-10 when using GNM. Again, varying parameters from the recommended 7\AA~ results in marginally better results, however no parametrization is able to reach the accuracy of mFRI.


\begin{figure}[H]
\begin{center}
\begin{tabular}{c}
\includegraphics[width=0.6\textwidth]{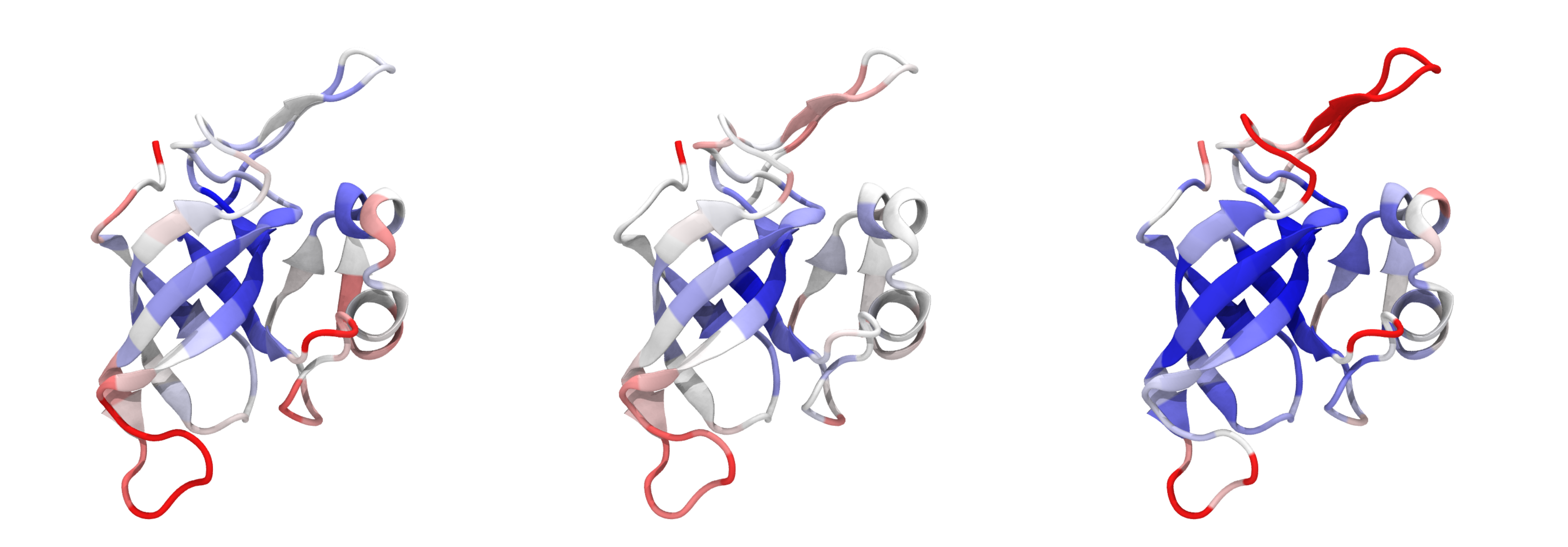}
\end{tabular}
\begin{tabular}{c}
\includegraphics[width=0.6\textwidth]{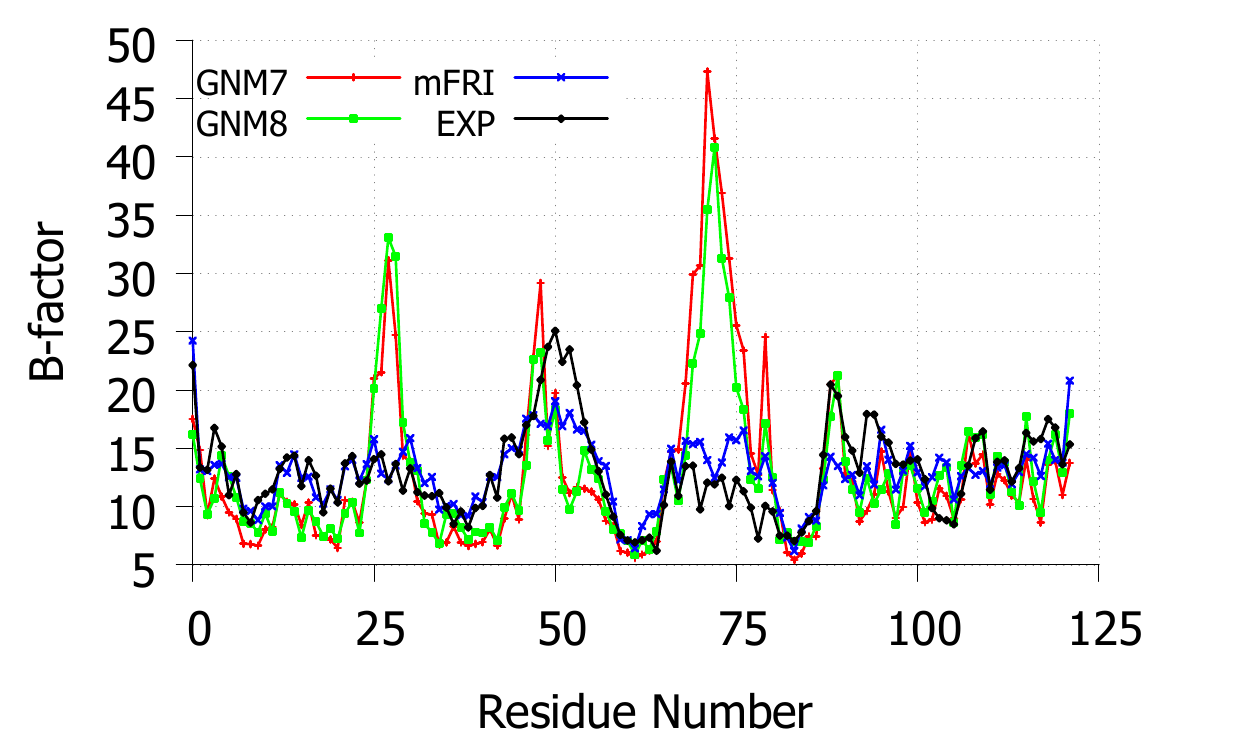}
\end{tabular}
\end{center}
\caption{ Top, a visual comparison of experimental B-factors (left), FRI predicted B-factors (middle) and GNM predicted B-factors (right) for the ribosomal protein L14 (PDB ID:1WHI). Bottom, The experimental and predicted B-factor values plotted per residue.
}
\label{1WHI}
\end{figure}

The final example is a biologically important molecule, ribosomal subunit L14, a component of the 60S ribosomal subunit\cite{Davies1996}. Depicted in Figure \ref{1WHI}, L14 is a structurally diverse protein containing regions of alpha helix, beta-barrel, parallel beta strands and a beta-hairpin motif. The pattern of flexibility predicted by GNM for this structure is shown to be over exaggerated, i.e., rigid areas are predicted to be more rigid than they actually are and vice verse. This pattern exists in most GNM results due to the use of a hard cutoff in the Kirchhoff matrix. Such a hard cutoff will inevitably lead to the overestimation of bond importance near the edge of the cutoff, therefore, if a large number of interactions exist for a particular atom near the cutoff point, there is likely to be a large error in the estimation of flexibility for that atom. This is likely what is happening with the errors in GNM calculation of the proteins in Figures \ref{2HQK}, \ref{1v70} and \ref{1WHI}, the protein at the end of the chain may be near the edge of the cutoff distance for many interactions with the bulk of the proteins. While adjusting GNM's cutoff distance may temper the error being introduced, it cannot eliminate it completely unless they change to a  soft-decaying kernel method such as FRI.
{Nevertheless, soft-decaying kernel based methods can only alleviate the problem. They   
do not deliver satisfactory B-factor predictions unless a multiscale strategy is employed.  We note that it is not obvious how to incorporate a multiscale strategy in matrix diagonalization based methods. }


\section*{Acknowledgments}

This work was supported in part by NSF grants IIS-1302285  and DMS-1160352, NIH Grant R01GM-090208, and
MSU Center for Mathematical Molecular Biosciences Initiative.
 

 \newpage

\centerline{\bf\Large  Supplementary Material} 

\section*{Flexibility-rigidity index}\label{sec:Flexibility}

In FRI, the topological connectivity of a biomolecule is measured by rigidity index and flexibility index. In particular, the rigidity index represents the protein density profile. Consider an  $N$-atom representation of a biomolecule. The coordinates of these atoms are given as $\{ {\bf r}_{j}| {\bf r}_{j}\in \mathbb{R}^{3}, j=1,2,\cdots, N\}$. We denote $  \|{\bf r}_i-{\bf r}_j\|$ the Euclidean space distance between the $i$th   atom  and the $j$th  atom. A general correlation kernel, $ \Phi( \|{\bf r} - {\bf r}_j \|;\eta_{j})$, is  a  real-valued monotonically decreasing function satisfying
\begin{eqnarray}\label{eq:couple_matrix1-1}
\Phi( \|{\bf r} - {\bf r}_j \|;\eta_{j})&=&1 \quad {\rm as }\quad  \|{\bf r} - {\bf r}_j \| \rightarrow 0\\\
\Phi( \|{\bf r} - {\bf r}_j \|;\eta_{j})&=&0 \quad {\rm as }\quad  \|{\bf r} - {\bf r}_j \| \rightarrow\infty,
\end{eqnarray}
where $\eta_{j}$ is an atomic type dependent scale parameter.
The correlation between the $i$th and $j$th particles is given by
\begin{eqnarray}\label{eq:couple_matrix0}
{C}_{ij} =  \Phi( \|{\bf r}_i - {\bf r}_j \|;\eta_{j}).
\end{eqnarray}
The correlation matrix $\{ C_{ij}\}$ can be computed to visualize the connectivity among protein particles.

We   define a position (${\bf r}$)  dependent rigidity function or density function  \cite{KLXia:2013d,Opron:2014}
\begin{eqnarray}\label{eq:rigidity3}
 \mu({\bf r}) & = & \sum_{j=1}^N w_{j} \Phi( \|{\bf r} - {\bf  r}_j \|;\eta_{j} ),
 \end{eqnarray}
 where $w_{j}$ is an atom type dependent weight. For example,  carbon,  nitrogen and  phosphorus atoms can have different weights. Although  Delta sequences of the positive type discussed in an earlier work  \cite{GWei:2000} are all good choices,   generalized exponential  functions
\begin{eqnarray}\label{eq:couple_matrix1}
\Phi(\|{\bf r} - {\bf r}_j \|;\eta_{j}) =    e^{-\left(\|{\bf r} - {\bf r}_j \|/\eta_{j}\right)^\kappa},    \quad \kappa >0
\end{eqnarray}
and  generalized Lorentz functions
\begin{eqnarray}\label{eq:couple_matrix2}
 \Phi(\|{\bf r} - {\bf r}_j \|;\eta_{j}) = \frac{1}{1+ \left( \|{\bf r} - {\bf r}_j \|/\eta_{j}\right)^{\upsilon}},  \quad  \upsilon >0
 \end{eqnarray}
have been commonly used in our recent work    \cite{KLXia:2013d,Opron:2014}.
Since the  rigidity function can be directly interpreted as a density function, it can been used to define the rigidity surface of a biomolecule by taking an isovalue.  By taking $\kappa =2$
in Eq. (\ref{eq:couple_matrix1}), we result in a formula for a Gaussian surface.

Similarly, we define a position (${\bf r}$)  dependent flexibility function  \cite{KLXia:2013d,Opron:2014}
\begin{eqnarray}\label{eq:flexibility1}
 F({\bf r}) & = & \frac{1}{\sum_{j=1}^N w_{j} \Phi( \|{\bf r} - {\bf  r}_j \|;\eta_{j} )}.
 \end{eqnarray}
This function is well defined in the computational domain containing the biomolecule. The flexibility function can be visualized by its projection on a given surface, such as the solvent excluded surface of a biomolecule.

The rigidity index for the $i$th particle is defined as
\begin{eqnarray}\label{eq:rigidity4}
 \mu_i & = & \sum_{j=1}^N w_{j} \Phi( \|{\bf r}_i - {\bf  r}_j \|;\eta_{j} ).
 \end{eqnarray}
Here $ \mu_i$ measures the total density or rigidity at the $i$th particle. In a similar manner, we define a set of  flexibility indices by
\begin{eqnarray}\label{eq:flexibility2}
 f_i & = & \frac{1}{\sum_{j=1}^N w_{j} \Phi( \|{\bf r}_i - {\bf  r}_j \|;\eta_{j} )}.
 \end{eqnarray}
 The flexibility index $ f_i$ is directly associated with  the  B-factor  of $i$th particle
\begin{eqnarray}\label{eq:regression}
 B_i^t = a f_i + b, \quad \forall i =1,2,\cdots,N
\end{eqnarray}
where $ \{B_i^t\}$ are  theoretically predicted B-factors,  and $a$ and $b$ are two  constants to be determined by a simple linear regression. This allows us to use  experimental  data to validate the FRI method. In our earlier work  \cite{KLXia:2013d,Opron:2014}, we set $w_j=1$ for the coarse-grained C$_\alpha$  representation of proteins. We have also developed parameter free  FRI (pfFRI), such as $(\kappa=1, \eta=3)$ and $(\upsilon=3, \eta=3)$, to make our FRI robust for protein  C$_\alpha$ B-factor prediction.

\subsection*{Multiscale FRI}\label{sec:MFRI}

The basic idea of multiscale FRI (mFRI) is quite simple. Since macromolecules are inherently multiscale in nature,
we utilize multiple correlation kernels that are parametrized at multiple scales to  characterize the multiscale  flexibility  of  macromolecules
\begin{eqnarray}\label{eq:flexibility3}
 f^{n}_i & = & \frac{1}{\sum_{j=1}^N w^{n}_{j} \Phi^{n}( \|{\bf r}_i - {\bf  r}_j \|;\eta^{n}_{j} )},
 \end{eqnarray}
where  $w^{n}_{j}$, $\Phi^{n}( \|{\bf r}_i - {\bf  r}_j \|;\eta^{n}_{j}) $ and $\eta^{n}_{j}$ are the corresponding quantities associated with the $n$th kernel.
We seek the minimization of the form
\begin{eqnarray}\label{eq:regression2}
{\rm Min}_{a^{n},b} \left\{ \sum_i \left| \sum_{n}a^n f^{n}_i + b-B^e_i\right|^2\right\}
\end{eqnarray}
where $\{B^e_i\}$ are the experimental B-factors. In principle, all parameters can be optimized. For simplicity and computational efficiency, we only determine $\{a^n\}$ and $b$ in the above  minimization process. For each kernel $\Phi^n$,  $w^n_j$  and $\eta^n_j$ will be selected according to the type of particles.

Specifically, for a simple C$_\alpha$ network, we can set  $w^n_j=1$ and choose a single kernel function parametrized at different scales. The predicted B-factors can be expressed as
\begin{eqnarray}\label{eq:flexibility4}
 B^{\rm mFRI}_i  = b+ \sum_{n=1}\frac{a^n}{\sum_{j=1}^N  \Phi( \|{\bf r}_i - {\bf  r}_j \|;\eta^{n} )}.
 \end{eqnarray}
The difference between Eqs. (\ref{eq:flexibility3}) and (\ref{eq:flexibility4}) is that, in Eqs. (\ref{eq:flexibility3}), both the kernel and the scale can be changed for difefrent $n$. In contrast, in Eq.  (\ref{eq:flexibility4}), only the scale is changed. One can use a given kernel, such as
\begin{eqnarray}\label{eq:couple_matrixn}
 \Phi(\|{\bf r} - {\bf r}_j \|;\eta^n) = \frac{1}{1+ \left( \|{\bf r} - {\bf r}_j \|/\eta^n\right)^{3}},
 \end{eqnarray}
to achieve good multiscale predictions.

\subsection*{Result evaluation}\label{sec:CC}

To quantitatively  assess the performance of the proposed multikernel based mFRI method, we  consider the correlation coefficient (CC)
\begin{eqnarray}\label{correlation}
   {\rm CC} =\frac{\eta^N_{i=1}\left(B^e_i-\bar{B}^e \right)\left( B^t_i-\bar{B}^t \right)}
   { \left[\eta^N_{i=1}(B^e_i- \bar{B}^e)^2\eta^N_{i=1}(B^t_i-\bar{B}^t)^2\right]^{1/2}},
\end{eqnarray}
where $\{B^t_i,  i=1,2,\cdots,N\}$ are a set of predicted B-factors by using the proposed method and $\{B^e_i, i=1,2,\cdots, N\}$ are a set of experimental B-factors extracted from the PDB file. Here $\bar{B}^t$ and $\bar{B}^e$ the statistical averages of theoretical and experimental B-factors, respectively.

\begin{figure}[H]
\begin{center}
\begin{tabular}{c}
\includegraphics[width=0.8\textwidth]{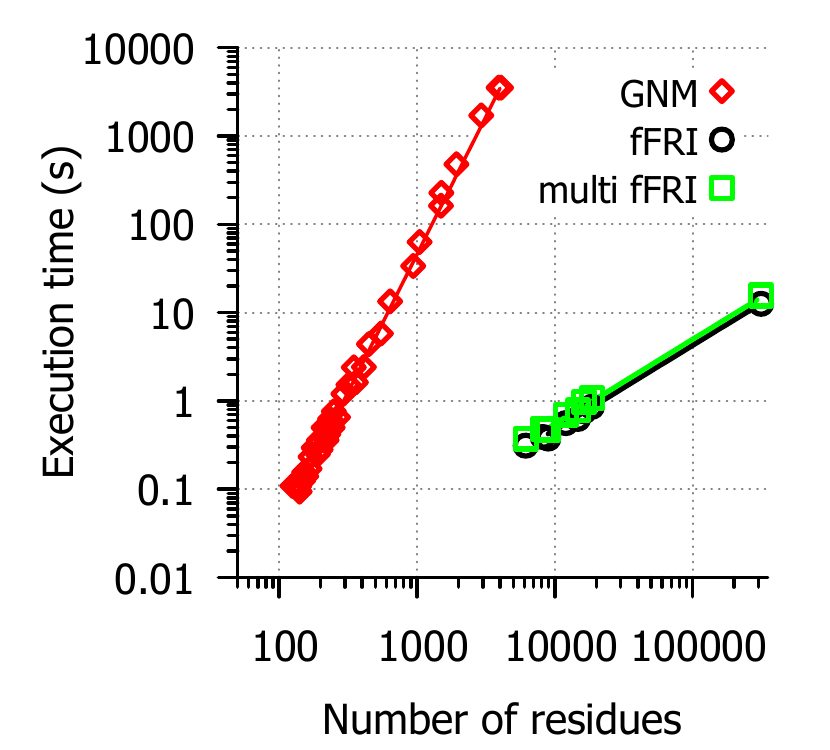}
\end{tabular}
\end{center}
\caption{Computational efficiency of multiscale fast FRI (multi fFRI) relative to single kernel fast FRI (fFRI) and  GNM. The data sets used for the present efficiency study are the same as those listed in Table VIII of Ref. \cite{Opron:2014}. The largest test molecule on the right is an HIV virus capsid, which has  than 313236 amino acid residues and takes about 30s for the mFRI method to compute its B-factor.
}
\label{complex}
\end{figure}

\subsection*{Parameterization in case studies}\label{sec:multisearch3}

\begin{table}[htbp]
  \centering
		\renewcommand\thetable{1}
	{	
	\scriptsize
\begin{tabular}{lcccccc}
\toprule
N & $\eta^2$=9\AA & $\eta^2$=12\AA& $\eta^2$=15\AA & $\eta^2$=17\AA & $\eta^2$=20\AA & $\eta^2$=25\AA \\
\midrule
0-99 & 0.055 & 0.083 & 0.100 & 0.102 & 0.097 & 0.083 \\
100-199 & 0.061 & 0.093 & 0.101 & 0.100 & 0.099 & 0.093 \\
200-299 & 0.051 & 0.087 & 0.097 & 0.097 & 0.095 & 0.087 \\
300-399 & 0.069 & 0.108 & 0.115 & 0.119 & 0.123 & 0.108 \\
400-499 & 0.079 & 0.126 & 0.148 & 0.157 & 0.155 & 0.126 \\
500+ & 0.064 & 0.107 & 0.136 & 0.143 & 0.140 & 0.107 \\
Overall & 0.060 & 0.094 & 0.106 & 0.108 & 0.106 & 0.094 \\
\bottomrule
\label{imp_table}
\end{tabular}
\caption{Improvements in mean  correlation coefficients  (MCC) for the B-factor prediction of a set of 364 proteins due to the introduction of an additional kernel parametrized at a large scale $(\eta^2)$. Two exponential kernels with $\kappa=25$ are employed. The first kernel's scale value is set to $\eta^1=7.0$\AA~ in all cases. The second kernel's scale value $(\eta^2)$  is varied and listed on the top of the table. Results are organized and split by the size of the structures based on the number of amino acids in order to show the impact of different  $\eta^2$ values on different sizes of proteins.}
  \label{long_table4}
	 }
\end{table}
In all of our case studies,   we have used both  mFRI and GNM to predict B-factors. When GNM performed poorly, different parameters were tried to see if there is a more ideal parametrization. The results of B-factor prediction are mapped on to the residues for visual comparison and shown plotted against the experimental values for more detail.
The mFRI method used in our case studies combines three kernels. After some testing we have decided upon using one kernel of exponential decay ($\kappa=1$) and two kernels of Lorentz type ($\upsilon=3$) with different scale ($\eta$) parameter values. {The choice of kernels and parameters is driven by the idea that each kernel should capture interactions of different ranges, e.g., short-, medium- and long-range interactions each being represented by a different kernel. The exponential kernel is chosen to represent the slowest decaying forces with $\eta^3=15$\AA~ and $\kappa=1$ while the two Lorentz type of kernels capture relative short- and medium-range interactions with parameters $\upsilon=3$, $\eta^1=3.0$\AA~ and $\eta^2=7$\AA, respectively.} The associated MCC for the 364 test set is 0.689, which is about 22\% better than what obtained by using the GNM method \cite{Opron:2014}. Other combinations of kernel parameters were tried in which the exponential kernel exhibited the quickest decay, however, they did not perform as well in B-factor prediction tests. The fast decaying Lorentz kernel, $\eta^1=3$\AA~ and $\upsilon=3$, may be well suited to capture the effect of chemical bonds due to its particular shape of decay which highly favors interactions below 3.0 \AA.

\subsection*{More detailed analysis of macromolecular multiscale behavior}
To further explore the importance of  multiscale methods, we use kernels having a sharp decay behavior similar to that of a Heaviside step function. This is achieved by setting $\kappa=25$ for the exponential type of correlation kernels.    In this case, the one-kernel FRI method behaves like the GNM method. The best performance for one-kernel FRI is obtained at $\eta^1=7$\AA ~ and the associated mean correlation coefficient  (MCC) for the 364 test set is 0.540, which is similar to that obtained by using GNM \cite{Opron:2014}. Obviously, the cutoff type of  kernel behavior obtained at $\kappa=25$ does not recognize any large-scale correlation beyond 7\AA~ in macromolecules. To capture large-scale correlations, we employ the second exponential kernel with its scale $(\eta^2>\eta^1)$ varying over a range of values as shown in Table  \ref{long_table4}.

To analyze the scale behavior due to protein size, we classify 364 proteins into 6 groups. The improvements of MCC  due to the introduction of an additional kernel are listed in Table  \ref{long_table4} for a number of large scale values $\eta^2$. First, the B-factor predictions from  all  size classes are significantly benefited from  the introduction of the large-scale kernel.  Additionally, at the scale value of  $\eta^2=17$\AA, the MCC is 0.648 and the associated improvement to the original FRI or GNM methods  is 20\% for the set of 364 proteins. Note that this multiscale improvement cannot be easily achieved by  GNM, NMA, or any other mode decomposition based methods. Moreover, the large-scale kernel leads to the most significant improvement in the B-factor prediction for relatively large proteins, i.e., proteins with 400-499  residues, which indicates that large proteins have more significant multiscale correlations than small proteins do.  Finally, the improvement in the B-factor prediction for proteins with more than 500 residues is not as much as that for   proteins with 400-499  residues, which indicates that two scales are not enough to capture all the multiscale correlations in proteins with more than 500 residues. This observation suggests that three or more scales are needed for the B-factor prediction of excessively large proteins.

\subsection*{Computational complexity  }\label{sec:comp}

It has been previously been demonstrated that the computational complexity of the single kernel FRI method is asymptotically of ${\cal O}(N^2)$. By making use of the cell lists algorithm, fFRI achieves a computational complexity of ${\cal O}(N)$. The addition of multiple kernels to the FRI method does not affect this aspect of scaling, however, the running time for B-factor prediction does increase with each additional kernel slightly. Indeed, the multi-kernel regression requires to  optimize one more parameter with the addition of each new kernel.  The impact of these changes on the running time of FRI based B-factor prediction is shown in Figure \ref{complex}. We employ the same data sets and test conditions as those described in our earlier paper \cite{Opron:2014} for the present test. The data used for testing mFRI and fFRI are the same as those used in testing the fFRI in Table VIII of Ref.  \cite{Opron:2014}. In  testing the GNM,  the same data set as that listed  in Table VIII of Ref.  \cite{Opron:2014} is employed.
\begin{figure}[H]
\begin{center}
\begin{tabular}{c}
\includegraphics[width=0.8\textwidth]{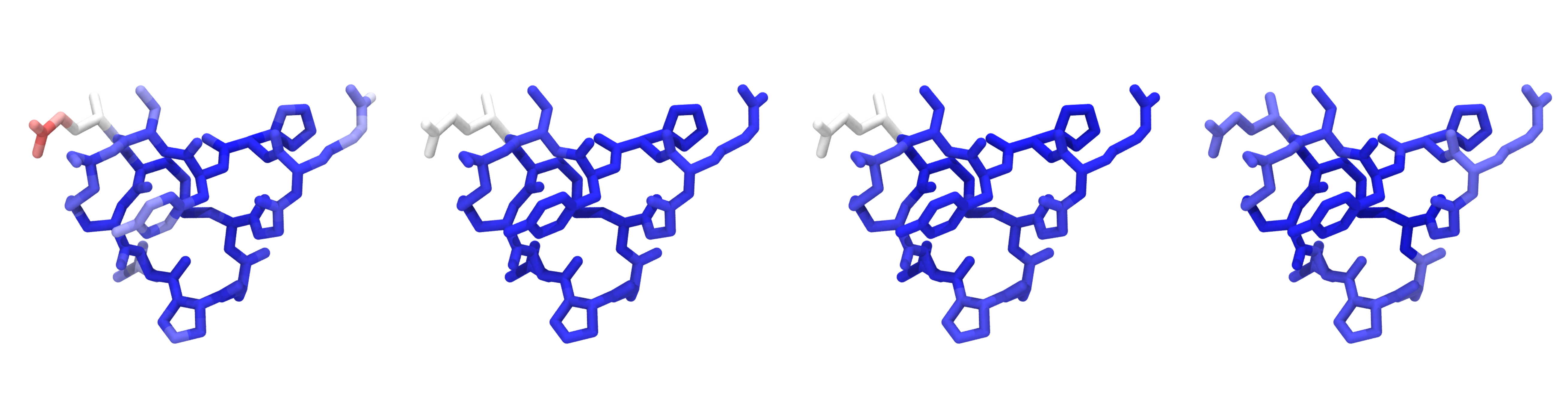}
\end{tabular}
\begin{tabular}{c}
\includegraphics[width=0.8\textwidth]{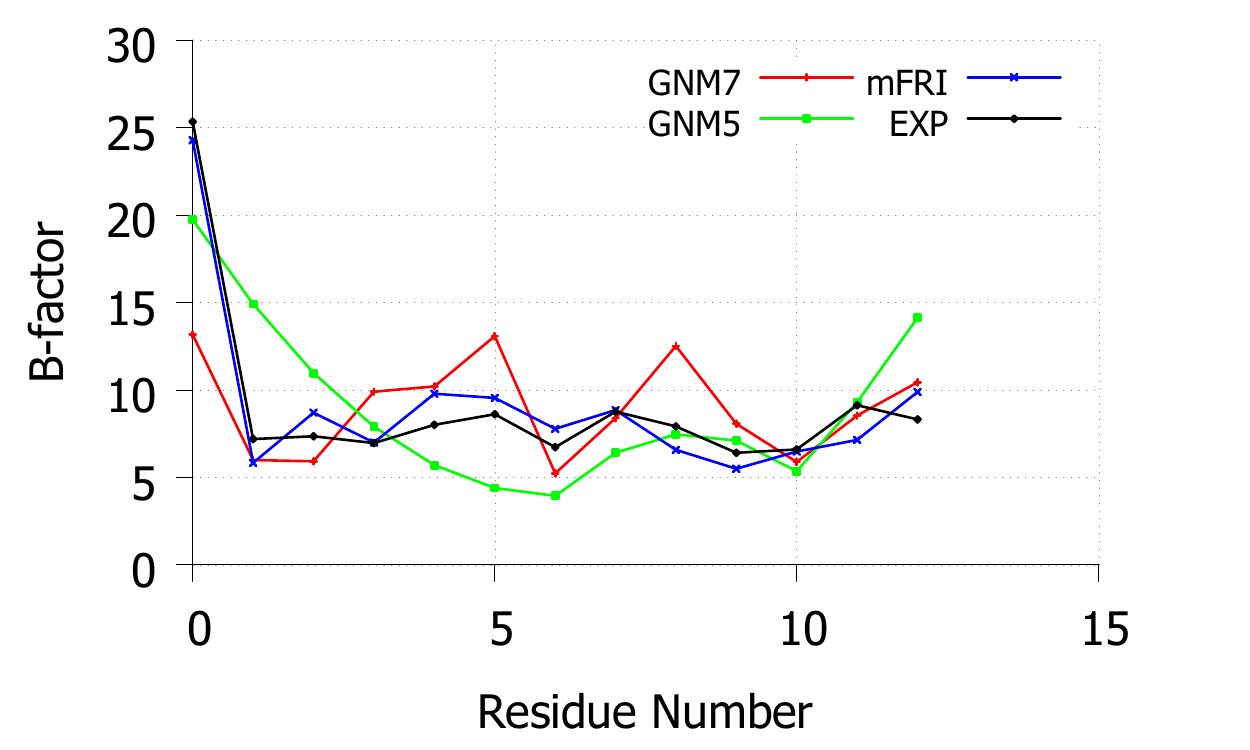}
\end{tabular}
\end{center}
\caption{ Top, a visual comparison of atomic experimental B-factors (far left), C-alpha experimental B-factors (left), mFRI predicted B-factors (middle) and GNM predicted B-factors (right) for the marine snail conotoxin (PDB ID:1NOT). Bottom, The experimental and predicted B-factor values plotted per residue.
}
\label{1NOT}
\end{figure}

Clearly the impact of extra kernels does not affect the essentially linear scaling of fFRI with lines of fit for fFRI and multikernel fast FRI (multi fFRI) being $t=7*10^{-6}*N^{0.957}$ and $t=8*10^{-6}*N^{0.959}$ respectively. The increase in computation time is minor especially for molecules with smaller numbers of atoms. In contrast, the line of fit for the GNM is $t=4*10^{-8}*N^{3.09}$ \cite{Opron:2014}. Note that each increase in one additional kernel leads to only one more fitting parameter, for which the fitting time is  negligibly small.  Only in extreme cases, with systems far larger than those currently studied atomistically, might single kernel FRI be preferred. Therefore, it is preferable to use multikernel based mFRI over single kernel FRI provided there is a significant increase in accuracy and reliability, as was demonstrated previously.  Note that the largest test molecule is an HIV virus capsid, which has  than 313236 amino acid residues. It would take the GNM more than 120 years to finish the prediction if the computer memory is not a problem. In contrast, the proposed mFRI does the job in about 30 seconds or less on a single workstation depending on the processing power.

\subsection*{Additional example}

\begin{figure}
\begin{center}
\begin{tabular}{c}
\includegraphics[width=0.8\textwidth]{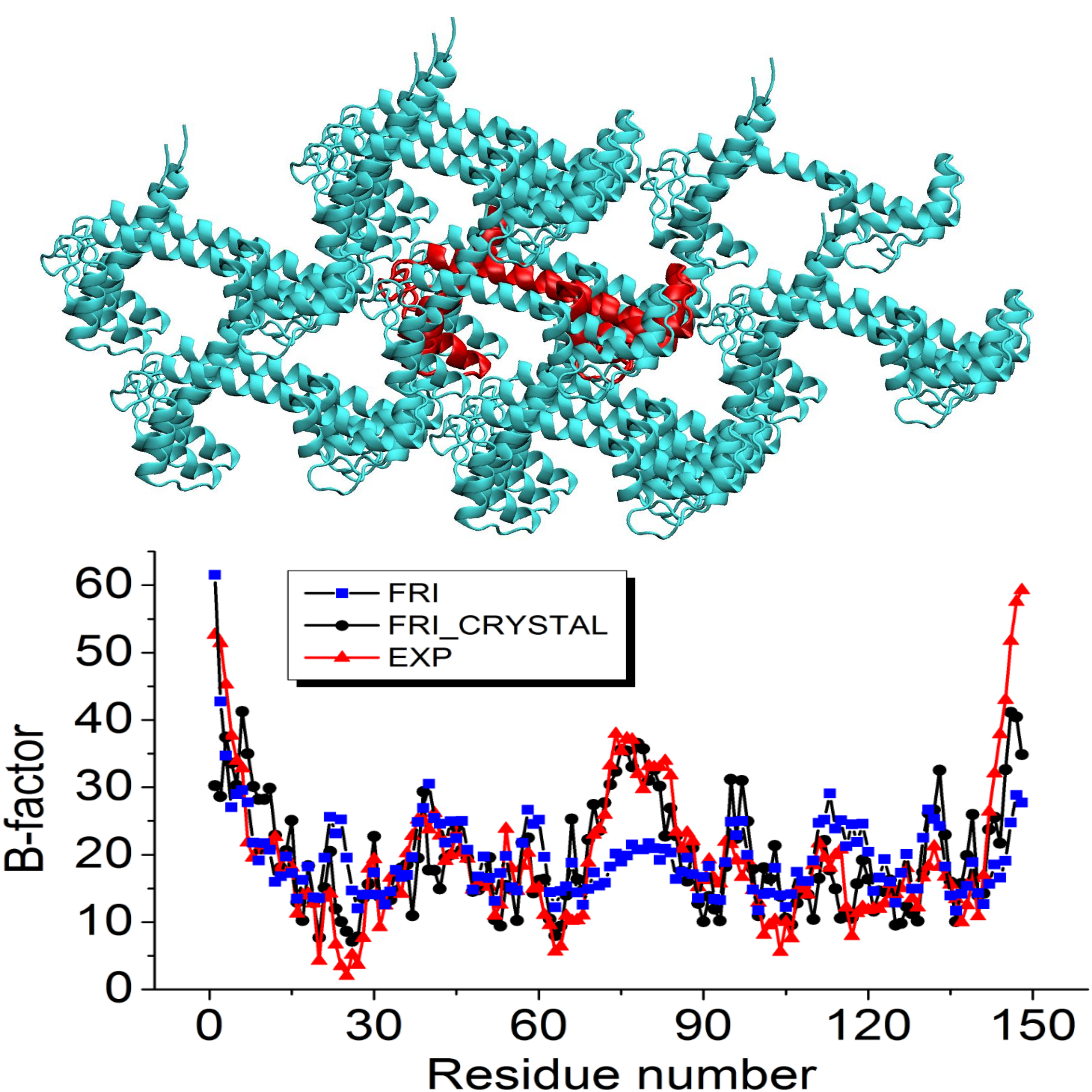}
\end{tabular}
\end{center}
\caption{The consideration of the crystal packing effect in FRI. 
Top: The crystal packing of 1OSA. The origin protein structure is shown in red color.
Bottom: The experimental B-factors and predicted ones. 
The correlation coefficient for FRI is 0.583. Using the packing information, the correlation coefficient value improves to 0.781.
}
\label{crystal}
\end{figure}
The proposed mFRI not only works better for the B-factor prediction of proteins, but also for  small molecules. One of such examples is a peptide molecule, a predatory marine snail toxin, shown in Figure \ref{1NOT}. This peptide adopts a cyclical secondary structure which is made up of two connected loops created by two disulfide bonds. In this structure there happens to be a particular residue at the beginning of the chain which is much more flexible than the others. This is a difficult case for flexibility prediction, especially coarse-grained predictions, as there may be side-chain interactions making large contributions to the flexibility of some atoms and there are two disulfide bonds that link. Nevertheless, mFRI is able to accurately reproduce the high flexibility of the first residue. GNM on the other hand is unable to recreate the pattern of flexibility at any parametrization. This is again due to the use of a hard cutoff in the GNM method and the use of a single kernel. The differences in distances between residues in this structure are too subtle to be captured by a method that treats distance with a hard cutoff. The kernels used in FRI are sensitive enough to detect the difference in distances between atoms in this structure which leads to finding the single stand-out residue.


\subsection*{ Crystal packing effect}
{The crystal packing is another very important element in theoretical B-factor prediction. The consideration of the crystal structure usually improves the accuracy of predictive models. We test our FRI model on a widely used example, i.e., protein 1OSA, which is another calmodulin structure. We consider all the copies in the crystal within 10 \AA~ distance of the protein. Figure \ref{crystal} demonstrates the crystal packing information. We modify our flexibility index by incorporation of all the $C_{\alpha}$ atoms within the above crystal packing structure. For our FRI model, we choose  the exponential kernel with $\kappa=1, \eta=4.0$\AA. The correlation coefficient for the original FRI is 0.583. When the crystal packing information is considered, the correlation coefficient value improves to 0.781. }

{
It is interesting to compare present Fig. \ref{crystal} with the Fig. 2 in our paper. Clear, two calmodulin structures have a similar behavior and are both difficult for FRI. This difficult can be  resolved either by a consideration of crystal effects or by using the proposed mFRI method. This example indicates that the proposed mFRI method can  take care some of the crystal parking effect.}

\end{document}